\documentclass[journal=nalefd,manuscript=article]{achemso}
\usepackage[version=3]{mhchem}
\usepackage{amssymb}

\author{Peter Rickhaus}
\author{Markus Weiss}
\email{Markus.Weiss@unibas.ch}
\author{Laurent Marot}
\author{Christian Sch\"onenberger}
\affiliation[unibas]{Department of Physics, University of Basel,
Klingelbergstrasse 82, 4056 Basel, Switzerland}

\title{Quantum Hall effect in graphene with super\-conducting electrodes}

\date{\today}

\begin{document}
\begin{abstract}
We have realized an integer quantum Hall system with
super\-conducting contacts by connecting graphene to niobium
electrodes. Below their upper critical field of 4 tesla, an integer quantum Hall effect coexists with
superconductivity in the leads, but with a plateau conductance that is larger than in the normal state. 
We ascribe this enhanced quantum Hall plateau conductance to Andreev processes at the graphene-superconductor
interface leading to the formation of so-called Andreev edge-states.
The enhancement depends strongly on the filling-factor, and is less pronounced
on the first plateau, due to the special nature of the zero energy Landau level in monolayer graphene.
\end{abstract}

\begin{figure}
  \includegraphics{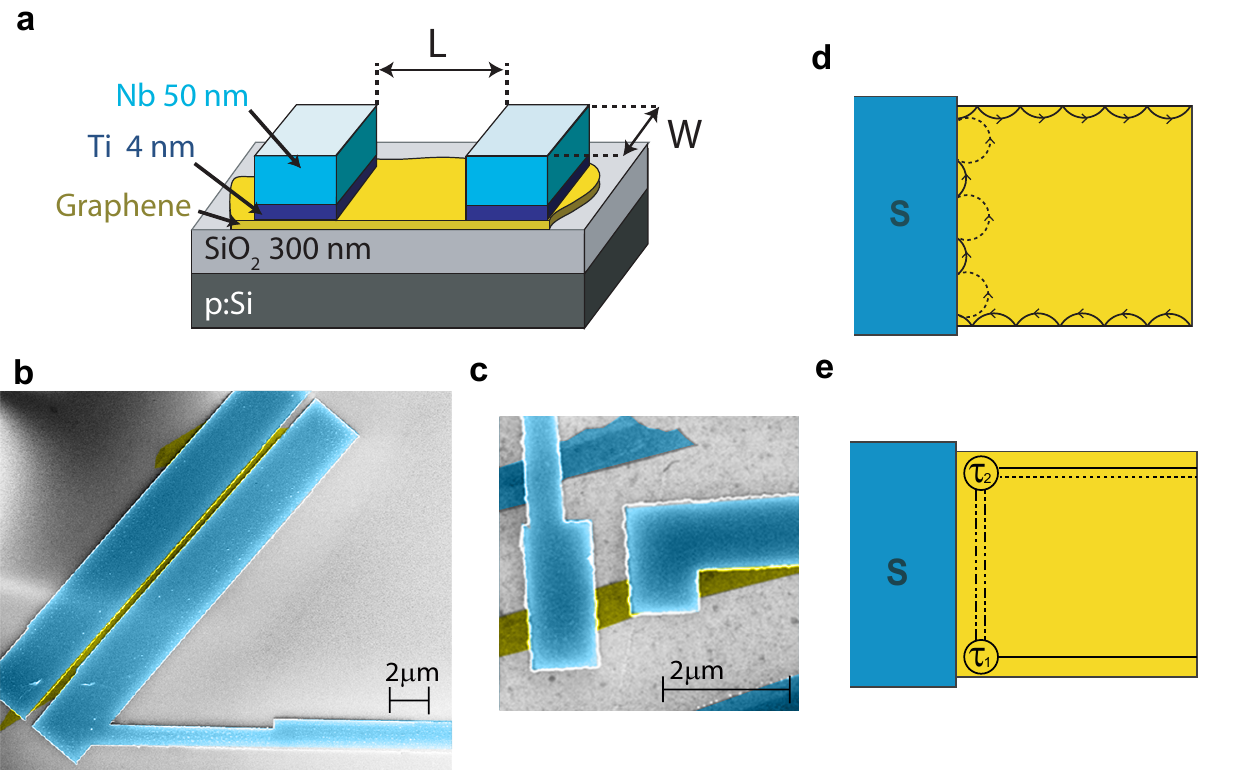}
  \caption{a) Schematic of sample structure with $L$ and $W$ denoting the length and width of the graphene device.
   b) false color SEM picture of a wide and c) a quadratic junction,
   of similar design as the ones measured. d) quasiclassical picture of an Andreev edge-state.
   e) Andreev edge state propagation in the Landauer B\"uttiker picture. Straight lines correspond to electrons, dotted lines to holes.}
  \label{figure1}
\end{figure}

The conductance of the interface between a two-dimensional electron gas (2DEG) and a superconductor (S) in a strong magnetic field
has received considerable interest in the past, both from experimental \cite{takayanagi1998,moore1999,eroms2005,batov2007}
and theoretical\cite{takagaki1998,asano2000,hoppe2000,chtchelkatchev2007,khaymovich2010} side. Experiments performed on InAs based 2DEG junctions with niobium
showed much stronger conductance oscillations than the usual Shubnikov-de Haas oscillations observed with normal (N) contacts\cite{eroms2005}.
Using niobium as a superconductor to contact an InAs based 2DEG, it has proven difficult to reach a regime of small
filling factor with clearly developed edge-states and the electrodes still in the super\-condcuting state. Using niobium nitride, which has
a higher critical field $B_{c2}$, only a very small conductance enhancement compared to the normal state could be observed~\cite{takayanagi1998}.
Similar to the zero-field case\cite{blonder1982}, theory predicts a doubling of conductance for a perfect 2DEG-S interface
in the quantum Hall regime\cite{hoppe2000}, although the electron trajectories are fundamentally different.
An electron hitting the 2DEG-S interface will, similar to the zero field case, perform an Andreev reflection,
forming a Cooper pair in the superconductor and retro\-reflecting a hole into the 2DEG.
As the retro\-reflected hole, which was created by Andreev reflection, lives in the same band as the impinging electron,
it has, in addition to carrying the opposite charge, also an effective mass of opposite sign.
As a consequence of this, it performs a cyclotron motion around the magnetic field
vector in the same sense as the electron (\ref{figure1}d). This gives rise to the formation of a so called
Andreev edge state \cite{hoppe2000}, that propagates along the 2DEG-S interface and consists, in a quasi-classical picture,
of alternating electron and hole orbits. For an interface with weak disorder and a small Fermi wavelength mismatch,
strong conductance oscillations as a function of magnetic field have been predicted due to interference between the electron
and hole parts of the Andreev edge-states\cite{hoppe2000,chtchelkatchev2007}.
At certain values of $B$ however, the maximum conductance of the ideal interface should still be reached.

In this letter we report on the realization of S-graphene-S devices based on niobium contacts.
Having a high upper critical magnetic field $B_{c2}$ of around $4$ T, the niobium contacts stay super\-conducting
when the graphene enters the quantum Hall effect regime, clearly evidenced by several quantum Hall plateaus
that are visible in a Landau level fan plot. We observe that the conductance in the plateau states is enhanced above the
quantized value in the normal state, and we argue that this enhancement is due to Andreev reflections at the graphene-S interface.

\begin{figure}
  \includegraphics{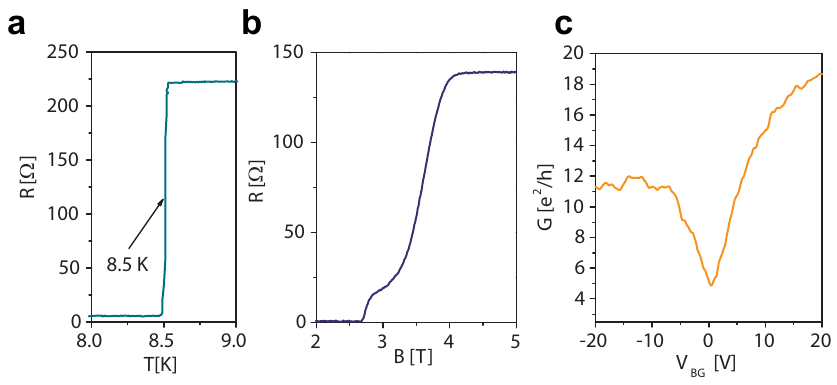}
  \caption{a) R(T) for a niobium teststrip. b) R(B) measured at $T=2.0$ K on the same teststrip with $B$ oriented perpendicular to the film plane.
c) backgate characteristics for one quadratic sample, measured at $T=20$ mK.}
  \label{figure2}
\end{figure}

\begin{figure}
  \includegraphics{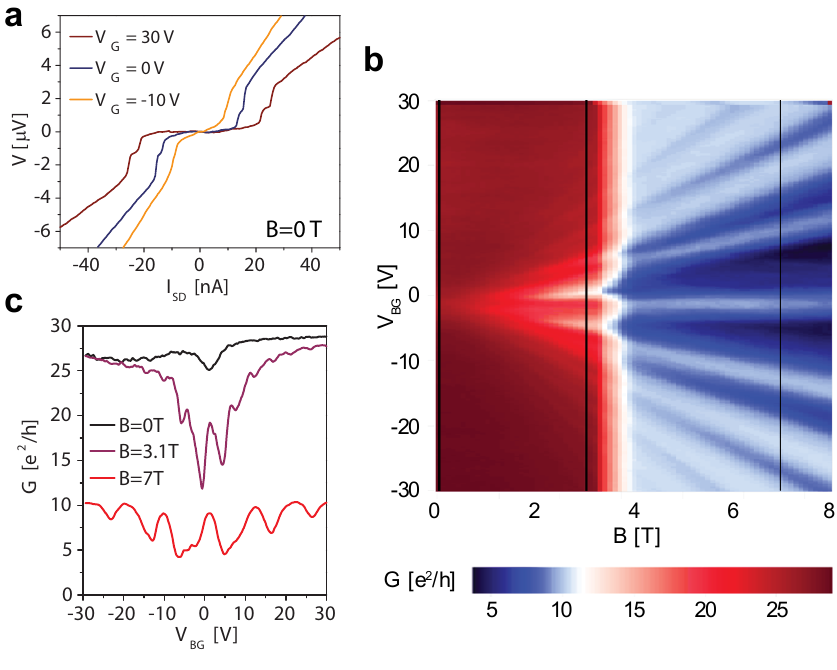}
  \caption{a) $V(I)$ curves measured on a wide sample at $T=20$ mK for three different backgate voltages.
b) Colorscale plot of differential conductance versus magnetic field and backgate voltage.
c) G(V$_{BG}$) for three different magnetic fields, as indicated in b).}
  \label{figure3}
\end{figure}

\begin{figure}
  \includegraphics{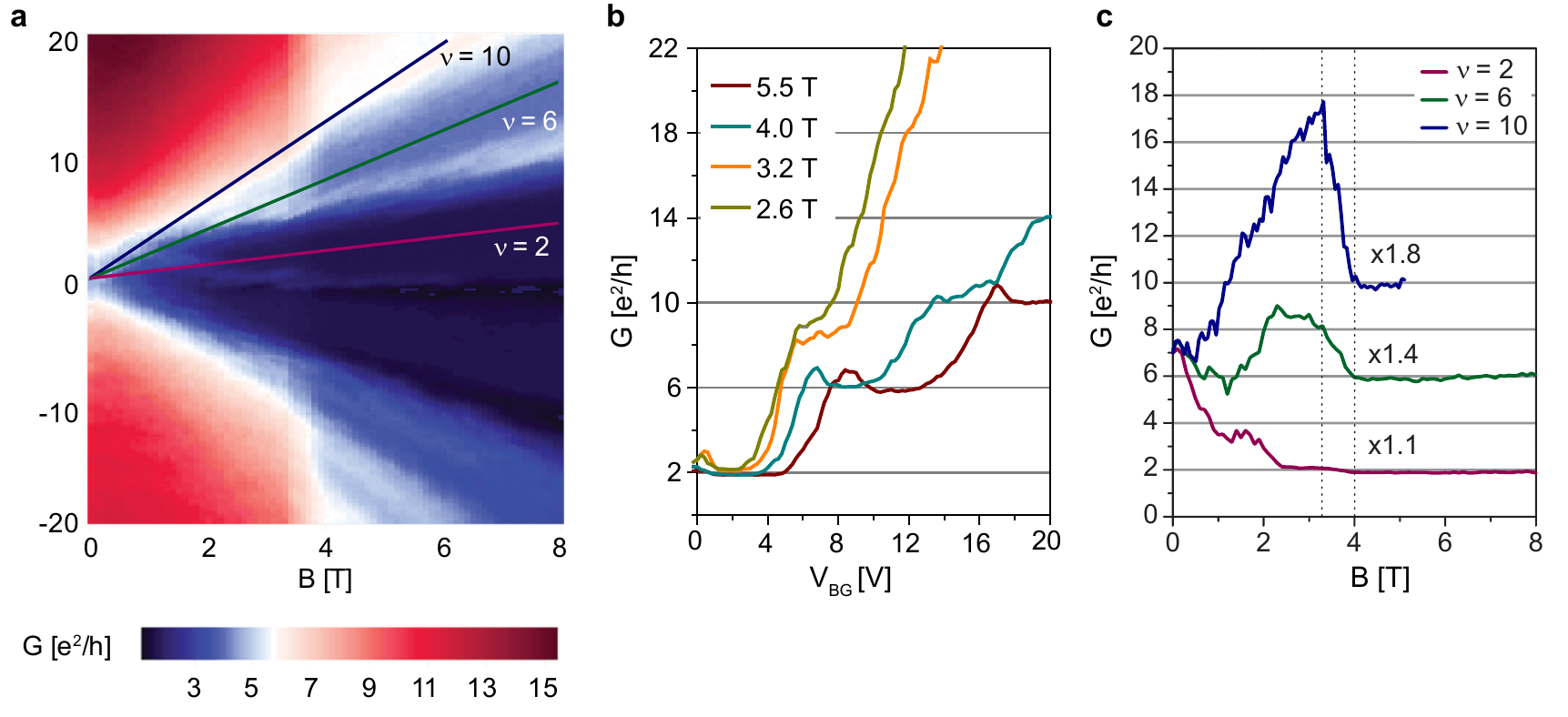}
  \caption{a) Colorscale plot of differential conductance versus backgate voltage and magnetic field for a quadratic sample. b) cuts through a)
taken at several constant magnetic fields for positive gate voltage. c) cuts through a) along the $ \nu=2$,$6$, and $10$ quantum Hall plateau,
as indicated by the tilted lines in a).}
\label{figure4}
\end{figure}

Monolayer graphene was prepared by mechanical exfoilation onto highly doped Si-substrates with $300$ nm of thermal oxide.
Appropriate flakes were located relative to predefined markers, the number of layers was determined by optical contrast,
and confirmed by confocal Raman spectroscopy. Electrodes were patterned by standard electron beam lithography using
a single layer of PMMA $950$K. A Ti($4$ nm) Nb($40$ nm) bilayer was deposited by magnetron sputtering (\ref{figure1}a).
We fabricated two types of samples: Wide samples with a junction width $W$ of
$5-32\mu$m and an electrode spacing $L \approx$ 400 nm (\ref{figure1}b),
and quadratic samples with W$\simeq$L and lateral dimensions of $\approx$2 $\mu$m (\ref{figure1}c).
Using a teststrip of $100$ $\mu$m length and $10$ $\mu$m width that was deposited together with the electrodes,
the transition temperature of the Ti/Nb bilayer was determined to be $T_c=8.5$ K (\ref{figure2}a), only slightly below
$9.25$ K, the bulk critical temperature of niobium. In a second test experiment, the upper critical magnetic field at $T=2.0$ K was determined
to be $B_{c2}=4$ T (\ref{figure2}b), for the magnetic field oriented perpendicular to the film plane.
The critical magnetic field showed small sample to sample variations, and increased slightly upon lowering the temperature from $2.0$ K
down to base temperature. To realize transparent contacts to graphene we found the thickness of the titanium contact layer to be crucial:
samples with less than $4$ nm of Ti showed exponentially increasing resistance for decreasing temperature.
Samples were cooled down in a dilution refrigerator with a base temperature of $T=20$ mK that was equipped with
a two stage filtering system, consisting of $\pi$-filters at room temperature and a dedicated high-frequency filter
at base temperature \cite{bluhm2008,spietz2006}.
For the Josephson current measurements shown in \ref{figure3}a an additional two-stage low-pass filter \cite{jarillo2006} close to the sample was used.
Two terminal conductance measurements on quadratic samples were corrected for a contact resistance that was determined by matching
the quantum Hall plateau conductances at $B > B_{c2}$ to the values expected for monolayer graphene\cite{abanin2008,williams2009}.
The contact resistance for the sample shown in \ref{figure4} was determined at B=5.5T, where superconductivity
in the leads is suppressed and plateaus at $\nu=$2, 6, and 10 are well developed. This normal state contact resistance was
then substracted from all measured conductance values, with B above and below the critical field of the electrodes.

The conductance as a function of backgate voltage $V_{BG}$ is shown in \ref{figure2}c for a quadratic sample.
The Dirac point appears close to zero gate voltage with a conductance value of $G\approx 5$ e$^2$/h.
We estimate the field effect mobility of this device to $\mu \approx 3000$ cm$^2$/Vs.
The quality of the super\-conducting contacts is confirmed by the measurement of a gate dependent Josephson current through the graphene at $B=0$,
shown in \ref{figure3}a. This measurement was performed on a short, wide sample with $L=400$ nm and $W=30$ $\mu$m.

The high critical field of the niobium contacts and the large cyclotron energy of monolayer graphene
\begin{equation}
E_N=sign(N) \sqrt{2 e \hbar v_F^2 |N|B}
\end{equation}
amounting to $30$ meV$\cdot \sqrt{|N| B[tesla]}$, allow to enter the quantum Hall regime while keeping the electrodes super\-conducting.
\ref{figure3}b shows the conductance of the wide sample measured as a function of gate voltage and magnetic field.
One can see in this figure that the conductance in the normal state above the critical field (right part)
is lower than in the super\-conducting state (left part). On the plot we also recognize a set of lines that are
caused by Landau level formation. These lines are most prominent in the normal state, but are also seen to extend into the super\-conducting region.
The observation of the quantum Hall effect in a two-terminal configuration is not
straightforward, as it is always complicated by a mixture of $\sigma_{xx}$ and $\sigma_{xy}$\cite{abanin2008}.
As the wide sample of \ref{figure3} had an aspect ratio $W/L$ of about $70$, the two-terminal conductance $G(B)$ is dominated
by $\sigma_{xx}$ and no flat quantum Hall plateaus are visible.
The magnetic field, where a Landau level is completely filled can however still be distinguished  by a pronounced minimum in $G(B)$.
The minima in $G(B)$, which are clearly visible at high magnetic fields in \ref{figure3}b and as well in cuts in  \ref{figure3}c
extend down to about $2$ T, where the electrodes are super\-conducting, giving a magnetic field range of more than $1$ T,
where clearly separated Landau levels coexists with superconductivity in the leads.

In two-terminal quantum Hall measurements, quadratic samples have the advantage of showing clear plateaus in the quantum Hall regime,
despite the mixing between $\sigma_{xx}$ and $\sigma_{xy}$.
Conductance as a function of V$_{BG}$ and $B$ measured on a quadratic sample is shown in \ref{figure4}a.
Comparable results were obtained on one other sample of similar design.
The data show a clear increase in conductance when lowering $B$ to below $4$ T (B$_{c2}$).
As in this case transport can happen via Andreev reflection processes \cite{andreev1964} an increase of $G$ is possible.
We quantify the conductance increase by evaluating the conductance ratio $G$($B=3.2$ T)/$G$($B=4$ T) taken along cuts
through \ref{figure4}a at constant filling factor $\nu$ (\ref{figure4}c), where
\begin{equation}
\nu = \frac {n h}{e B} = \frac{C_{BG}\cdot(V_{BG}-V_{CNP})}{e} \cdot \frac{h}{e B}
\end{equation}
with the gate capacitance $C_{BG}$, and $V_{CNP}$ the position of the Dirac point on the gate voltage axis.
The two field values have been chosen because below the upper magnetic field of $B=4$ T (taken to be $B_{c2}$) the conductance starts to deviate from the
quantized value and at $B=3.2$ T the resistance of the teststrip (see \ref{figure2}b) has decreased to less than half of the normal state value.
The niobium electrode is in the mixed state at this magnetic field, with magnetic flux penetrating the film in the form of flux vortices.
At $B=3.2$ T the vortices are sufficiently diluted, so that they do not affect Andreev processes at the graphene-S interface.
Looking at the conductance change only in a narrow field range, we can also exclude
a significant contribution due to the overlap between neighboring Landau levels.
Going from $\nu=2$ to $\nu=10$, we see that the conductance ratio increases from a factor of $1.1$ ($\nu=2$) over $1.4$ ($\nu=6$) to $1.8$ ($\nu=10$).
The conductance increase is more pronounced when a
larger number of quantum Hall edge-states is involved in transport. Note that the total increase in $G$ stays well below a factor of two,
which would be the limit given by Andreev reflection on a fully transparent S-N interface, as predicted by BTK theory \cite{blonder1982}.

The conductance of a conventional 2DEG-S interface in the quantum Hall effect regime depends on several factors, including Fermi-wavelength mismatch
between 2DEG and superconductor, interface barrier strength, and edge disorder.
For weak disorder the scattering at a super\-conducting contact can be described in an
effective Landauer B\"uttiker picture \cite{hoppe2000,chtchelkatchev2007,khaymovich2010},
as indicated in \ref{figure1}e. An incoming electron edge-state can scatter into two Andreev-edge states
with probabilities $\tau_1$ and $1-\tau_1$, respectively. The Andreev edge states propagate along the 2DEG-S interface,
and scatter themselves to either an electron or a hole edge-state at the opposite edge of the sample with probabilities $\tau_2$
and $1-\tau_2$, respectively. The total conductance is then given by\cite{khaymovich2010}
\begin{equation}
G_{NS}=2G_0\left[\tau_1\left(1-\tau_2\right)+\tau_2\left(1-\tau_1\right)\right],
\end{equation}
with $G_0=2$ e$^2$/h. Depending on $\tau_1$ and $\tau_2$, the conductance of a system with one spin-degenerate edge-state
populated can take on any value between $0$ and $4$ e$^2$/h.
For strong disorder, on the other hand, Andreev edge-states are destroyed and the conductance of a 2DEG-N
interface is recovered\cite{chtchelkatchev2007}.

In contrast to a conventional 2DEG, where the Andreev reflected hole is retro\-reflected, Andreev reflection in graphene can
be specular when the Fermi energy is close to the Dirac point \cite{beenakker2006}. As in this case the hole is back\-reflected
into a different band, its effective mass has the same sign as the electron effective mass. Due to its positive charge,
it now performs cyclotron motion in the opposite sense as the electron.
As the phase of the hole wave\-function in then also opposite to the one of the electron, the Andreev edge-state that corresponds to the
zero energy Landau level is not affected by interference between electron and hole orbits as in the case of Andreev retro\-reflection.
With only the zero energy Landau level populated ($\nu=2$), the conductance of the graphene-S interface in the quantum Hall regime
only depends on the angle $\theta$ between the valley polarizations of incoming and outgoing edge-state\cite{akhmerov2007}:
\begin{equation}
G_{NS}=\frac{2 e^2}{h}(1-\cos{\theta}),
\end{equation}
where $\cos{\theta}={\bf\hat{\nu_1}}\cdot{\bf \hat{\nu_2}}$, and ${\bf \hat{\nu_{1/2}}}$ are the valley polarizations
of the incoming/outgoing edge-states, respectively. In our situation, incoming and outgoing egde-state run on opposite
sides of the sample in opposite direction. If we assume identical edges on opposite sides, we get
opposite valley polarizations (${\bf \hat{\nu_1}}=-{\bf \hat{\nu_2}}$) for in- and outgoing edge-states,
which would lead to a conductance of $4$ e$^2$/h. According to Ref. \cite{akhmerov2007}, deviations from this value must then
be due to intervalley scattering. Turning to the measurements presented in \ref{figure4},
we see that the conductance on the first quantum Hall plateau at $\nu=2$ increases by a factor of $1.1$ when the magnetic field is reduced to
below the critical field of the electrodes. The conductance of the first edge-state is therefore only slightly enhanced
above the normal state value of $2$ e$^2$/h, which according to Ref. \cite{akhmerov2007} must be due to strong inter\-valley scattering.
The enhancement of the total conductance by a factor of $1.4$ ($\nu$=6) and $1.8$ ($\nu$=10) when the second and third
edge-state are filled, respectively, is consistent with the second
and third edge-state both contributing a doubled conductance of $8$ e$^2$/h to the total conductance,
while the conductance of the first edge-state remains roughly constant at $2$ e$^2$/h.
Contrary to $N=0$, the $N=1$ and $N=2$ Landau levels form valley degenerate edge-states\cite{brey2006,abanin2006}, whose propagation should
not be as sensitive to the structure of the graphene edges as in the case of the zero energy Landau level edge-state.
In addition they are located further away from the disordered sample edges, and probably suffer less from
scattering than the outer edge-state, and therefore show a stronger conductance enhancement due to Andreev reflections
than the $N=0$ edge-state.

In conclusion we have demonstrated monolayer graphene with super\-conducting contacts made from niobium.
The contacts show a high critical magnetic field, that allows to enter the quantum Hall effect regime of graphene, while keeping the
contacts super\-conducting.
The super\-conducting proximity effect manifests itself at zero magnetic field in the form of a Josephson current,
and as a marked conductance enhancement in the plateau states of the quantum Hall effect at higher magnetic fields.
The conductance increase depends on the number of edge-states involved in transport.
Whereas the conductance of the first edge-state is almost unaffected by the superconductivity in the leads,
the total conductance increases substantially when the second and third edge-states are populated.
We think that this is due to the special nature of the zero-energy Landau level in monolayer graphene,
which is composed of both electron and hole states, but is not valley degenerate on the edge.
The propagation of the $N=0$ edge-state will therefore strongly depend on the structure of the graphene edge,
and will be strongly modified in the presence of inter\-valley scattering.
For clean edges, we would expect a doubling of conductance for our sample geometry,
which apparently is suppressed by strong intervalley scattering, caused by a strongly disordered graphene edge.
The edge-states that originate from higher Landau levels are valley degenerate and should be less sensitive to the structure
of the edge and to disorder, giving a much stronger conductance enhancement due to Andreev processes than for the N=0 edge-state.

\acknowledgement

We thank Roland Steiner for help in sputter deposition, and Carlo Beenakker, Romain Maurand, and
Frank Freitag for helpful discussions.
Financial support by the Swiss NCCR on Quantum Science and Technology
is gratefully acknowledged.


\begin{mcitethebibliography}{21}
\providecommand*\natexlab[1]{#1}
\providecommand*\mciteSetBstSublistMode[1]{}
\providecommand*\mciteSetBstMaxWidthForm[2]{}
\providecommand*\mciteBstWouldAddEndPuncttrue
  {\def\EndOfBibitem{\unskip.}}
\providecommand*\mciteBstWouldAddEndPunctfalse
  {\let\EndOfBibitem\relax}
\providecommand*\mciteSetBstMidEndSepPunct[3]{}
\providecommand*\mciteSetBstSublistLabelBeginEnd[3]{}
\providecommand*\EndOfBibitem{}
\mciteSetBstSublistMode{f}
\mciteSetBstMaxWidthForm{subitem}{(\alph{mcitesubitemcount})}
\mciteSetBstSublistLabelBeginEnd
  {\mcitemaxwidthsubitemform\space}
  {\relax}
  {\relax}

\bibitem[Takayanagi and Akazaki(1998)Takayanagi, and Akazaki]{takayanagi1998}
Takayanagi,~H.; Akazaki,~T. \emph{Physica B} \textbf{1998}, \emph{249-251},
  462\relax
\mciteBstWouldAddEndPuncttrue
\mciteSetBstMidEndSepPunct{\mcitedefaultmidpunct}
{\mcitedefaultendpunct}{\mcitedefaultseppunct}\relax
\EndOfBibitem
\bibitem[Moore and Williams(1999)Moore, and Williams]{moore1999}
Moore,~T.~D.; Williams,~D.~A. \emph{Physical Review B} \textbf{1999},
  \emph{59}, 7308\relax
\mciteBstWouldAddEndPuncttrue
\mciteSetBstMidEndSepPunct{\mcitedefaultmidpunct}
{\mcitedefaultendpunct}{\mcitedefaultseppunct}\relax
\EndOfBibitem
\bibitem[Eroms et~al.(2005)Eroms, Weiss, Boeck, Borghs, and
  Z{\"u}licke]{eroms2005}
Eroms,~J.; Weiss,~D.; Boeck,~J.~D.; Borghs,~G.; Z{\"u}licke,~U. \emph{Physical
  Review Letters} \textbf{2005}, \emph{95}, 107001\relax
\mciteBstWouldAddEndPuncttrue
\mciteSetBstMidEndSepPunct{\mcitedefaultmidpunct}
{\mcitedefaultendpunct}{\mcitedefaultseppunct}\relax
\EndOfBibitem
\bibitem[Batov et~al.(2007)Batov, Sch{\"a}pers, Chtchelkatchev, Hardtdegen, and
  Ustinov]{batov2007}
Batov,~I.~E.; Sch{\"a}pers,~T.; Chtchelkatchev,~N.; Hardtdegen,~H.;
  Ustinov,~A.~V. \emph{Physical Review B} \textbf{2007}, \emph{76},
  115313\relax
\mciteBstWouldAddEndPuncttrue
\mciteSetBstMidEndSepPunct{\mcitedefaultmidpunct}
{\mcitedefaultendpunct}{\mcitedefaultseppunct}\relax
\EndOfBibitem
\bibitem[Takagaki(1998)]{takagaki1998}
Takagaki,~Y. \emph{Physical Review B} \textbf{1998}, \emph{57}, 4009\relax
\mciteBstWouldAddEndPuncttrue
\mciteSetBstMidEndSepPunct{\mcitedefaultmidpunct}
{\mcitedefaultendpunct}{\mcitedefaultseppunct}\relax
\EndOfBibitem
\bibitem[Asano(2000)]{asano2000}
Asano,~Y. \emph{Physical Review B} \textbf{2000}, \emph{61}, 1732\relax
\mciteBstWouldAddEndPuncttrue
\mciteSetBstMidEndSepPunct{\mcitedefaultmidpunct}
{\mcitedefaultendpunct}{\mcitedefaultseppunct}\relax
\EndOfBibitem
\bibitem[Hoppe et~al.(2000)Hoppe, Z{\"u}licke, and Sch{\"o}n]{hoppe2000}
Hoppe,~H.; Z{\"u}licke,~U.; Sch{\"o}n,~G. \emph{Physical Review Letters}
  \textbf{2000}, \emph{84}, 1804\relax
\mciteBstWouldAddEndPuncttrue
\mciteSetBstMidEndSepPunct{\mcitedefaultmidpunct}
{\mcitedefaultendpunct}{\mcitedefaultseppunct}\relax
\EndOfBibitem
\bibitem[Chtchelkatchev and Burmistrov(2007)Chtchelkatchev, and
  Burmistrov]{chtchelkatchev2007}
Chtchelkatchev,~N.~M.; Burmistrov,~I.~S. \emph{Physical Review B}
  \textbf{2007}, \emph{75}, 214510\relax
\mciteBstWouldAddEndPuncttrue
\mciteSetBstMidEndSepPunct{\mcitedefaultmidpunct}
{\mcitedefaultendpunct}{\mcitedefaultseppunct}\relax
\EndOfBibitem
\bibitem[Khaymovich et~al.(2010)Khaymovich, Chtchelkatchov, Shereshevskii, and
  Melnikov]{khaymovich2010}
Khaymovich,~I.~M.; Chtchelkatchov,~N.~M.; Shereshevskii,~I.~A.; Melnikov,~A.~S.
  \emph{Europhysics Letters} \textbf{2010}, \emph{91}, 17005\relax
\mciteBstWouldAddEndPuncttrue
\mciteSetBstMidEndSepPunct{\mcitedefaultmidpunct}
{\mcitedefaultendpunct}{\mcitedefaultseppunct}\relax
\EndOfBibitem
\bibitem[Blonder et~al.(1982)Blonder, Tinkham, and Klapwijk]{blonder1982}
Blonder,~G.~E.; Tinkham,~M.; Klapwijk,~T.~M. \emph{Physical Review B}
  \textbf{1982}, \emph{25}, 4515\relax
\mciteBstWouldAddEndPuncttrue
\mciteSetBstMidEndSepPunct{\mcitedefaultmidpunct}
{\mcitedefaultendpunct}{\mcitedefaultseppunct}\relax
\EndOfBibitem
\bibitem[Bluhm and Moler(2008)Bluhm, and Moler]{bluhm2008}
Bluhm,~H.; Moler,~K.~A. \emph{Review of Scientific Instruments} \textbf{2008},
  \emph{79}, 014703\relax
\mciteBstWouldAddEndPuncttrue
\mciteSetBstMidEndSepPunct{\mcitedefaultmidpunct}
{\mcitedefaultendpunct}{\mcitedefaultseppunct}\relax
\EndOfBibitem
\bibitem[Spietz et~al.()Spietz, Teufel, and Schoelkopf]{spietz2006}
Spietz,~L.; Teufel,~J.; Schoelkopf,~R.~J. \emph{unpublished
  (arXiv:cond-mat/0601316v1)} \relax
\mciteBstWouldAddEndPunctfalse
\mciteSetBstMidEndSepPunct{\mcitedefaultmidpunct}
{}{\mcitedefaultseppunct}\relax
\EndOfBibitem
\bibitem[Jarillo-Herrero et~al.(2006)Jarillo-Herrero, van Dam, and
  Kouwenhoven]{jarillo2006}
Jarillo-Herrero,~P.; van Dam,~J.~A.; Kouwenhoven,~L.~P. \emph{Nature}
  \textbf{2006}, \emph{439}, 953\relax
\mciteBstWouldAddEndPuncttrue
\mciteSetBstMidEndSepPunct{\mcitedefaultmidpunct}
{\mcitedefaultendpunct}{\mcitedefaultseppunct}\relax
\EndOfBibitem
\bibitem[Abanin and Levitov(2008)Abanin, and Levitov]{abanin2008}
Abanin,~D.; Levitov,~L.~S. \emph{Physical Review B} \textbf{2008}, \emph{78},
  035416\relax
\mciteBstWouldAddEndPuncttrue
\mciteSetBstMidEndSepPunct{\mcitedefaultmidpunct}
{\mcitedefaultendpunct}{\mcitedefaultseppunct}\relax
\EndOfBibitem
\bibitem[Williams et~al.(2009)Williams, Abanin, DiCarlo, Levitov, and
  Marcus]{williams2009}
Williams,~J.~R.; Abanin,~D.~A.; DiCarlo,~L.; Levitov,~L.~S.; Marcus,~C.~M.
  \emph{Physical Review B} \textbf{2009}, \emph{80}, 045408\relax
\mciteBstWouldAddEndPuncttrue
\mciteSetBstMidEndSepPunct{\mcitedefaultmidpunct}
{\mcitedefaultendpunct}{\mcitedefaultseppunct}\relax
\EndOfBibitem
\bibitem[Andreev(1964)]{andreev1964}
Andreev,~A.~F. \emph{Soviet Physics JETP} \textbf{1964}, \emph{19}, 1228\relax
\mciteBstWouldAddEndPuncttrue
\mciteSetBstMidEndSepPunct{\mcitedefaultmidpunct}
{\mcitedefaultendpunct}{\mcitedefaultseppunct}\relax
\EndOfBibitem
\bibitem[Beenakker(2006)]{beenakker2006}
Beenakker,~C. W.~J. \emph{Physical Review Letters} \textbf{2006}, \emph{97},
  067007\relax
\mciteBstWouldAddEndPuncttrue
\mciteSetBstMidEndSepPunct{\mcitedefaultmidpunct}
{\mcitedefaultendpunct}{\mcitedefaultseppunct}\relax
\EndOfBibitem
\bibitem[Akhmerov and Beenakker(2007)Akhmerov, and Beenakker]{akhmerov2007}
Akhmerov,~A.~R.; Beenakker,~C. W.~J. \emph{Physical Review Letters}
  \textbf{2007}, \emph{98}, 157003\relax
\mciteBstWouldAddEndPuncttrue
\mciteSetBstMidEndSepPunct{\mcitedefaultmidpunct}
{\mcitedefaultendpunct}{\mcitedefaultseppunct}\relax
\EndOfBibitem
\bibitem[Brey and Fertig(2006)Brey, and Fertig]{brey2006}
Brey,~L.; Fertig,~H.~A. \emph{Physical Review B} \textbf{2006}, \emph{73},
  195408\relax
\mciteBstWouldAddEndPuncttrue
\mciteSetBstMidEndSepPunct{\mcitedefaultmidpunct}
{\mcitedefaultendpunct}{\mcitedefaultseppunct}\relax
\EndOfBibitem
\bibitem[Abanin et~al.(2006)Abanin, Lee, and Levitov]{abanin2006}
Abanin,~D.~A.; Lee,~P.~A.; Levitov,~L. \emph{Physical Review Letters}
  \textbf{2006}, \emph{96}, 176803\relax
\mciteBstWouldAddEndPuncttrue
\mciteSetBstMidEndSepPunct{\mcitedefaultmidpunct}
{\mcitedefaultendpunct}{\mcitedefaultseppunct}\relax
\EndOfBibitem
\end{mcitethebibliography}
\providecommand*\mcitethebibliography{\thebibliography}
\csname @ifundefined\endcsname{endmcitethebibliography}
  {\let\endmcitethebibliography\endthebibliography}{}

\end{document}